\documentclass{ifacconf}
\usepackage{amssymb}
\usepackage{graphicx} 
\usepackage{natbib}
\usepackage{amsmath} 
\usepackage{bm}      
\usepackage{booktabs}
\usepackage{algorithm}
\usepackage{algpseudocode}
\usepackage{placeins}
\usepackage{comment}
\begin{document}
\begin{frontmatter}

\title{Ensemble-Based Modelling of Optogenetic Bioprocesses under Phototoxicity}

\thanks[footnoteinfo]{\,SER e-mail: sebastian.espinelrios@ucd.ie.}

\author[First]{Yufan Liu} 
\author[Second]{Sam Ferrer} 
\author[First]{Sebastián Espinel-Ríos} 

\address[First]{School of Chemical and Bioprocess Engineering,
University College Dublin, Ireland}
\address[Second]{Toulouse Graduate School of Chemical, Materials, and Industrial Engineering, France}

\begin{abstract}
Optogenetics enables methanol-free protein production in the yeast \textit{Komagataella phaffii} using light as a control input, facilitated by the blue-light-responsive transcription factor EL222. This opens the door to more programmable protein production in the food and biopharmaceutical industries. However, excessive illumination can inhibit growth and protein accumulation in some of these engineered cell factories. Previous hybrid modelling used black-box regression across sparsely sampled conditions to infer light-dependent kinetic parameters, without explicitly representing intracellular regulation. Here, we develop a multiscale mechanistic model linking EL222 activation, transcription, and translation to protein, biomass, and substrate dynamics. A latent light-associated stress state represents photoinhibitory effects on growth, transcription, and translation. An ensemble-based parameter-estimation strategy combines repeated global--local optimisation with screening for parameter sets supporting locally stable intracellular steady states under specified light conditions. Using available batch data for yeast-enhanced green fluorescent protein production in a \textit{K. phaffii} strain carrying eight EL222 gene copies, the ensemble-mean predictions captured the main experimental trends in protein accumulation and growth across six constant light intensities. Parameter variability was propagated to time-course predictions, providing an empirical assessment of parameter-related uncertainty. Wider prediction bands coincided with less complete experimental coverage of the growth phases. Overall, the framework provides a mechanistic basis for future model-based optimisation and control of optogenetic bioprocesses under phototoxicity.
\end{abstract}


\end{frontmatter}

\section{Introduction}

Microbial protein production is a versatile biotechnological platform enabling the recombinant synthesis of enzymes, peptides, and antibodies for applications across the food and biopharmaceutical industries \citep{barone2023,hoffman2025}. Within this context, the yeast \textit{Komagataella phaffii} (formerly \textit{Pichia pastoris}) is widely used as a \textit{workhorse} for recombinant protein production, given its capacity for high-cell-density growth and eukaryotic post-translational modifications \citep{barone2023}. \textit{K. phaffii} bioprocesses generally employ a two-stage strategy based on the methanol-inducible alcohol oxidase 1 promoter, $\mathrm{P_{AOX1}}$. During the initial growth phase, cells are commonly cultivated on glycerol or glucose, supporting biomass accumulation while recombinant protein expression remains repressed. Following depletion of the initial carbon source, the production phase is initiated by feeding methanol, which serves as both a carbon source and an inducer \citep{barone2023}. This phase separation provides a form of dynamic metabolic control aimed at balancing biomass formation and recombinant protein synthesis.

Optogenetic regulation offers a methanol-free alternative in which light modulates gene expression, offering a more tunable metabolic response \citep{hoffman2022,hoffman2025}. Recently, \cite{hoffman2025} demonstrated blue-light-induced production of recombinant proteins in \textit{K. phaffii} using EL222, a light-responsive transcription factor. They experimentally investigated how EL222 gene copy number and light intensity jointly influenced growth and protein-production dynamics. Notably, increasing EL222 gene copy number enhanced sensitivity to light, a potential advantage in large-scale bioproduction where light penetration can be limiting. However, excessive illumination impaired growth and protein production in those high-copy-number cell factories, effects collectively referred to here as \textit{phototoxic effects}. The authors hypothesised that these effects arose from the generation of damaging reactive oxygen species (ROS) upon light activation of EL222.

To describe these interactions, \cite{hoffman2025} developed a hybrid model combining macroscopic balances for biomass, substrate, and intracellular protein with Gaussian-process predictions of kinetic parameters as functions of EL222 gene copy number and light intensity. This approach captured the observed growth and protein-production dynamics across the experimentally tested conditions. However, the effects of light and EL222 copy number were represented through data-driven kinetic parameters inferred from sparsely sampled experimental conditions rather than explicit intracellular regulatory states. Consequently, the model does not resolve the mechanistic activation, transcriptional, and phototoxic stress-related processes governing the observed bioprocess dynamics. In addition, its Gaussian-process-predicted parameter functions exhibited pronounced local variations and substantial predictive uncertainty in unsampled regions, raising concerns about model generalisation and interpretability.

In this work, we develop a multiscale mechanistic model that links light-mediated intracellular regulation to macroscopic bioprocess behaviour. The formulation is intended for optogenetic protein-production systems rate-limited by phototoxicity, as can occur at high EL222 gene copy numbers. To this end, we link active EL222 and recombinant messenger ribonucleic acid (mRNA) to protein, biomass, and substrate dynamics, with a latent light-associated stress state accounting for phototoxic inhibition. We demonstrate the modelling framework using experimental data from a glucose-grown \textit{K. phaffii} strain engineered for optogenetic production of yeast-enhanced green fluorescent protein (yEGFP) and carrying eight copies of the EL222 gene. This was the highest copy number examined in the light-intensity characterisation experiments reported by \cite{hoffman2025}, and the strain's susceptibility to phototoxicity makes it a suitable case study for the proposed framework.

While model formulation draws on known or hypothesised mechanisms, parametrisation presents an additional challenge. Parameter values may be unavailable from the literature or may not be directly transferable to the strain and operating conditions of interest. Although parameters can be estimated from experimental data, the available measurements and experimental conditions may be insufficient to uniquely determine them, allowing different parameter combinations to reproduce the observed dynamics. Biological models can exhibit both structural and practical nonidentifiability \citep{villaverde2019identifiability,wieland_structural_2021}. Structural nonidentifiability arises when distinct parameter sets produce indistinguishable model outputs, whereas practical nonidentifiability reflects insufficient information in the available data to estimate parameters with adequate precision. Assessing parameter uncertainty and propagating it through the dynamics therefore enables the associated uncertainty to be estimated in the time domain.

Building on previous ensemble-based modelling approaches \citep{massonis_improving_2023,huber_model-based_2024,huber_combining_2026}, we perform several independent parameter-estimation runs to generate an ensemble of parameter sets. This is done following a hybrid global--local optimisation approach. We additionally apply steady-state stability analysis to retain parameter sets that admit biologically achievable intracellular steady states under specified constant illumination. Variability across these parameter sets provides an ensemble-based estimation of parameter uncertainty. Simulating the model with each retained parameter set yields trajectories whose spread provides a corresponding estimation of dynamic prediction uncertainty.

\section{Mathematical model formulation}

Extending the optogenetic gene-expression and growth-dependent translation structures of \cite{benisch_optogenetic_2023}, we introduce a dynamic latent stress state to represent photoinhibitory effects. This state links light exposure to inhibition of growth, transcription and translation. The mechanistic model is defined by the state vector $\bm{x}=[a,R,m,P,X,S]^\mathsf{T}$, where $a \in [0,1]$ is the active fraction of the intracellular EL222 transcription factor, $R \in \mathbb{R}_{\geq 0}$ represents a latent intracellular light-associated stress response, and $m \in \mathbb{R}_{\geq 0}$ represents intracellular mRNA produced by the light-inducible expression system. The state $P \in \mathbb{R}_{\geq 0}$ represents intracellular yEGFP concentration in arbitrary units, quantified as fluorescence normalised by OD$_{600}$ (optical density at 600 nm), $X \in \mathbb{R}_{\geq 0}$ represents cell density expressed as OD$_{600}$, and $S \in \mathbb{R}_{\geq 0}$ is the extracellular glucose concentration in $\mathrm{g}\,\mathrm{L}^{-1}$. The external input is the blue light intensity $I \in \mathbb{R}_{\geq 0}$, expressed in $\mu\mathrm{mol}\,\mathrm{m}^{-2}\,\mathrm{s}^{-1}$.

The inhibitory effects associated with the stress state $R$ are represented by Hill functions:
\begin{equation}
\phi_i(R)
=
\frac{K_{R,i}^{\alpha_i}}
     {K_{R,i}^{\alpha_i}+R^{\alpha_i}},
\qquad i \in \{X,m,P\},
\label{eq:stress_inhibition}
\end{equation}
where $K_{R,i}$ is the stress half-saturation constant, and $\alpha_i$ determines the steepness of the response. The functions $\phi_X$, $\phi_m$, and $\phi_P$ thus describe inhibition of growth, transcription, and translation, respectively.

The effective specific growth rate follows Monod-type kinetics, modified by stress-dependent inhibition:
\begin{equation}
\mu(S,R)
=
\mu_{\max}\frac{S}{K_S+S}\phi_X(R),
\label{eq:specific_growth_rate}
\end{equation}
where $\mu_{\max}$ is the maximum specific growth rate and $K_S$ is the substrate half-saturation constant. For brevity, the explicit dependence of $\mu$ on $S$ and $R$ is omitted below.

Following the optogenetic activation structure used by \citet{benisch_optogenetic_2023}, EL222 activation is driven by blue light and reversed by first-order inactivation:
\begin{equation}
\frac{\mathrm{d}a}{\mathrm{d}t}
=
k_{\mathrm{on}}I(1-a)-k_{\mathrm{off}}a,
\label{eq:el222_activation}
\end{equation}
where $k_{\mathrm{on}}$ determines light-dependent activation, with $(1-a)$ being the fraction of transcription factor available for activation, and $k_{\mathrm{off}}$ is the dark-state reversion rate constant. The active fraction is mapped to an effective active transcription-factor abundance through $TF=\eta_E a$, where $\eta_E$ sets the transcription-factor abundance scale.

The mRNA balance includes basal transcription, transcription factor-dependent activation, degradation, and dilution by growth. We retain the Hill-type activation structure of \cite{benisch_optogenetic_2023} and extend it with stress-dependent inhibition of transcription:
\begin{equation}
\frac{\mathrm{d}m}{\mathrm{d}t}
=
\left(
k_{m,\mathrm{basal}}
+
k_m\frac{TF^n}{K_m^n+TF^n}
\right)\phi_m(R)
-
(k_{m,d}+\mu)m,
\label{eq:mrna_dynamics}
\end{equation}
where $k_{m,\mathrm{basal}}$ is the basal transcription rate, $k_m$ is the maximum inducible transcription rate, $K_m$ is the transcription-factor half-saturation constant, and $n$ is the Hill exponent. The loss terms account for first-order mRNA degradation with rate constant $k_{m,d}$ and dilution at the growth rate $\mu$. 

Intracellular yEGFP production is proportional to mRNA abundance and scaled by growth-dependent translational capacity. Following the structure used by \cite{benisch_optogenetic_2023}, yet incorporating additional stress-dependent inhibition, the protein balance is:
\begin{equation}
\frac{\mathrm{d}P}{\mathrm{d}t}
=
k_{\mathrm{trans}}
\frac{\mu}{\mu_{\max}}
m\phi_P(R)
-
(k_{P,d}+\mu)P,
\label{eq:protein_dynamics}
\end{equation}
where $k_{\mathrm{trans}}$ is the translation-rate coefficient and $k_{P,d}$ is the first-order protein degradation rate constant. The ratio $\mu/\mu_{\max}$ represents the dependence of translational capacity on growth, reflecting the assumed relationship between growth rate and ribosomal abundance.

For batch cultivation, biomass accumulation and glucose depletion are described by:
\begin{align}
\frac{\mathrm{d}X}{\mathrm{d}t}
&= \mu X,
\label{eq:biomass_dynamics}\\
\frac{\mathrm{d}S}{\mathrm{d}t}
&= -\frac{\mu}{Y_{X/S}} X,
\label{eq:substrate_dynamics}
\end{align}
where $Y_{X/S}$ is the biomass yield on glucose.

Finally, the dynamics of the latent stress state, which is linked to all photoinhibitory functions, include basal and illumination-dependent generation, together with first-order clearance and dilution by growth:
\begin{equation}
\frac{\mathrm{d}R}{\mathrm{d}t}
=
k_{R,\mathrm{basal}}
+
k_R I^{p}
-
(k_{R,\mathrm{c}}+\mu)R,
\label{eq:stress_dynamics}
\end{equation}
where $k_{R,\mathrm{basal}}$ is the basal stress-generation rate, $k_R$ scales the illumination-dependent contribution, $p$ determines its dependence on light intensity, and $k_{R,\mathrm{c}}$ is the stress-clearance rate constant.

\section{Ensemble-based parameter estimation}

The constant model parameters are collected in the vector $\bm{\theta}\in\mathbb{R}_{\geq 0}^{23}$. Parameter estimation was formulated as the minimisation of the weighted sum of squared differences between model predictions and experimental measurements:
\begin{equation}
\label{eq:fit_objective}
J(\bm{\theta})
=
\sum_{e\in\mathcal{E}}
\sum_{r\in\{P,X\}}
\sum_{k\in\mathcal{K}_{e,r}}
w_{e,r}
\left[
y_{e,r}^{\mathrm{mod}}(t_{e,k};\bm{\theta})
-
y_{e,r}^{\mathrm{exp}}(t_{e,k})
\right]^2.
\end{equation}
Here, $\mathcal{E}$ denotes the experimental conditions, each corresponding to a different constant light intensity. $\mathcal{K}_{e,r}$ contains the indices of sampling times at which output state $r$ was measured within experiment $e$; thus $t_{e,k}$ denotes the corresponding measurement time. The quantities $y_{e,r}^{\mathrm{exp}}$ and $y_{e,r}^{\mathrm{mod}}$ denote the experimental observations and corresponding model predictions, respectively. Time-course measurements were available for $P$ and $X$, while initial values were available for $P$, $X$, and $S$. The unmeasured initial states $a(0)$, $R(0)$, and $m(0)$ were computed analytically from the intracellular steady-state equations under assumed dark pre-culture conditions ($I=0$), with the pre-culture growth rate approximated by $\mu_{\max}$. These initial states were therefore determined by the model parameters rather than estimated independently.

The weights followed a mean-square scheme:
\begin{equation}
\label{eq:fit_weights}
M_{e,r}
=
\frac{1}{n_{e,r}}
\sum_{k\in\mathcal{K}_{e,r}}
\left[y_{e,r}^{\mathrm{exp}}(t_{e,k})\right]^2,
\qquad
\widetilde{w}_{e,r} = \frac{1}{M_{e,r}},
\end{equation}
where $n_{e,r}=|\mathcal{K}_{e,r}|$. These weights were subsequently divided by the largest weight within each experiment, yielding the normalised weights $w_{e,r}$ used in Eq. \eqref{eq:fit_objective}.

The bounded optimisation problem was:
\begin{equation}
\label{eq:fit_run}
\min_{\bm{\theta}_{\min}\leq\bm{\theta}\leq\bm{\theta}_{\max}}
J(\bm{\theta}).
\end{equation}
For each independent estimation run $i$, the initial parameter values $\theta_{0,j}^{(i)}$ were sampled from log-uniform distributions within  prescribed positive bounds:
\begin{equation}
\label{eq:random_start}
\begin{aligned}
z_j^{(i)}
&\sim
\mathcal{U}
\left(
\log_{10}\theta_{j,\min},
\log_{10}\theta_{j,\max}
\right),
\\
\theta_{0,j}^{(i)}
&=10^{z_j^{(i)}},
\qquad j=1,\ldots,23,
\end{aligned}
\end{equation}
where $\theta_{j,\min}$ and $\theta_{j,\max}$ are the lower and upper bounds of parameter $j$.

We used a hybrid global--local optimisation approach to combine exploration of the parameter space with local refinement~\citep{villaverde2019optimization}. In our implementation, each estimation run $i$ comprised two rounds of particle swarm optimisation (PSO)~\citep{kennedy1995}, followed by three rounds of Hooke--Jeeves optimisation~\citep{hooke1961}. PSO is a population-based global optimisation algorithm, whereas Hooke--Jeeves is a derivative-free local direct-search method. The parameter estimate returned by each round was used to initialise the next within the run $i$. 

The complete estimation procedure was repeated independently $N=500$ times, yielding the following candidate parameter ensemble:
\begin{equation}
\label{eq:candidate_ensemble}
\bm{\Theta}
=
\left\{
\hat{\bm{\theta}}^{(i)}
\right\}_{i=1}^{N},
\end{equation}
where $\hat{\bm{\theta}}^{(i)}$ denotes the numerical parameter estimate obtained from the $i$-th independent run.
Each ensemble member therefore represents a fitted parameterisation obtained under the same experimental data,
model structure, and parameter bounds.

Parameter estimation was implemented in Python using BasiCO \citep{bergmann_basico_2023}, which provides an interface to COPASI: a COmplex PAthway SImulator \citep{hoops_copasicomplex_2006}.

\section{Steady-state stability screening}

To assess whether the fitted models admit locally asymptotically stable intracellular steady states under reference illumination conditions, we screened the parameter sets in $\bm{\Theta}$ at fixed light intensities: $I\in\{0,5,10,30,50,70\}$
$\mu\mathrm{mol}\,\mathrm{m}^{-2}\,\mathrm{s}^{-1}$, matching the experimental conditions reported by \cite{hoffman2025}. Following a similar rationale to \cite{greene_acceleration_2017}, here stability was used as a biological plausibility criterion, reflecting the assumption that cells can maintain stable intracellular states under constant illumination and non-limiting substrate availability.

For this analysis, the dynamic equations for $X$ and $S$ were excluded, and the substrate-dependent growth expression in (\ref{eq:specific_growth_rate}) was replaced by:
\begin{equation}
\label{eq:ss_mu}
\mu(R)=\mu_{\max}\phi_X(R).
\end{equation}
This assumes non-limiting substrate availability while retaining stress-dependent growth inhibition.

Steady-state calculations and linear stability analysis were performed through BasiCO, which follows the general framework described below. For each fitted parameter set $\hat{\bm{\theta}}^{(i)}$ and light condition $I^{(e)}$, the steady state of the intracellular subsystem $\bm{x}_{\mathrm{int}}=[a,R,m,P]^\mathsf{T}$ satisfies:
\begin{equation}
\label{eq:ss_condition}
\bm{f}_{\mathrm{int}}\!\left(
\bm{x}_{\mathrm{int}}^{ss,(i,e)};
I^{(e)},\hat{\bm{\theta}}^{(i)}
\right)=\bm{0},
\end{equation}
where $\bm{f}_{\mathrm{int}}$ denotes the right-hand side of the intracellular model given Eq. \eqref{eq:ss_mu}.

For each finite steady-state solution obtained numerically, local linearisation is characterised by the Jacobian:
\begin{equation}
\label{eq:ss_jacobian}
\mathbf{A}^{(i,e)}
=
\left.
\frac{\partial\bm{f}_{\mathrm{int}}}
{\partial\bm{x}_{\mathrm{int}}}
\right|_{
\bm{x}_{\mathrm{int}}=\bm{x}_{\mathrm{int}}^{ss,(i,e)}
}.
\end{equation}

The eigenvalues satisfy:
\begin{equation}
\det\!\left(
\mathbf{A}^{(i,e)}-\lambda\mathbf{I}_4
\right)=0,
\end{equation}
where $\mathbf{I}_4$ is the identity matrix.
A computed steady state was classified as locally
asymptotically stable if:
\begin{equation}
\label{eq:ss_stability}
\alpha^{(i,e)}
=
\max_{j=1,\ldots,4}
\operatorname{Re}\!\left(\lambda_j^{(i,e)}\right)<0.
\end{equation}

A fitted parameter set was retained only if a steady state was found, as per Eq. \eqref{eq:ss_condition}, and Eq. \eqref{eq:ss_stability} was further satisfied under all six light conditions. The retained parameter sets form the actual ensemble used for parameter-distribution and trajectory-uncertainty analysis. The complete workflow is summarised in Algorithm~\ref{alg:ensemble_screening}.

\begin{algorithm}[t]
\footnotesize
\caption{Parameter estimation and screening}
\label{alg:ensemble_screening}
\begin{algorithmic}[1]
\For{$i = 1,\dots,N$}
    \State Sample $\bm{\theta}_0^{(i)}$ using (\ref{eq:random_start})
    \State Apply two global-search rounds followed by three local-search rounds
    \State Save the fitted set $\hat{\bm{\theta}}^{(i)}$
\EndFor
\State Initialize the retained ensemble $\mathcal{E}_{\mathrm{stable}}\gets\emptyset$
\For{each fitted model $\hat{\bm{\theta}}^{(i)}$}
    \State Construct the intracellular subsystem by excluding $X$ and $S$
       and setting Eq. \eqref{eq:ss_mu}
    \State Set $\mathrm{accepted}\gets\mathrm{true}$
    \For{each $I\in\{0,5,10,30,50,70\}$}
        \State Solve the steady-state condition in Eq. \eqref{eq:ss_condition}
        \If{no steady state is found numerically or the computed steady state has negative concentrations}
            \State Set $\mathrm{accepted}\gets\mathrm{false}$ and \textbf{break}
        \EndIf
        \State Compute the Jacobian in Eq. \eqref{eq:ss_jacobian} and its eigenvalues
        \If{the stability criterion (\ref{eq:ss_stability}) is not satisfied}
            \State Set $\mathrm{accepted}\gets\mathrm{false}$ and \textbf{break}
        \EndIf
    \EndFor
    \If{$\mathrm{accepted}$}
        \State $\mathcal{E}_{\mathrm{stable}}\gets
        \mathcal{E}_{\mathrm{stable}}\cup\{\hat{\bm{\theta}}^{(i)}\}$
    \EndIf
\EndFor
\end{algorithmic}
\end{algorithm}

\section{Parameter-Distribution and Trajectory-Uncertainty Analysis}

Table~\ref{tab:ce8_screening} summarises the results of Algorithm~\ref{alg:ensemble_screening}. Among the 500 candidate parameter sets obtained through random-start parameter estimation, 124 produced intracellular steady states with strictly negative real parts for all four Jacobian eigenvalues under all six light conditions. Of the rejected sets, 205 were excluded because the steady-state calculation did not converge under at least one condition, leaving stability undetermined. Notably, no steady states with negative concentrations were reported. All steady states admitted to stability screening yielded four finite Jacobian eigenvalues, allowing the stability criterion to be evaluated in every case. The remaining 171 rejected parameter sets had a non-negative maximum Jacobian eigenvalue real part at a computed steady state under at least one light condition and therefore failed the stability criterion in (\ref{eq:ss_stability}).

\begin{table}[t]
\centering
\caption{Summary of the ensemble screening workflow.}
\label{tab:ce8_screening}
\begin{tabular}{p{0.80\columnwidth}l}\toprule

\textbf{Outcome}& \textbf{Count}\\\midrule

Initial candidate parameter sets& $500$\\

Rejected because the steady-state calculation did not converge under at least one light condition & $205$\\

Rejected upon not satisfying (\ref{eq:ss_stability}) following successful steady-state calculations& $171$\\

Retained after satisfying (\ref{eq:ss_stability}) under all six light conditions &$124$\\ \bottomrule
\end{tabular}
\end{table}   

The screening thus supplemented parameter estimation with an additional selection criterion, retaining multiple fitted parameter combinations that supported locally asymptotically stable intracellular steady states under all reference light conditions. Model selection therefore incorporated both the assumed biological stability requirement and numerical steady-state feasibility. However, numerical non-convergence left the stability of 205 parameter sets undetermined, representing a substantial fraction of the candidate ensemble. Adjustments to solver settings might resolve some of these cases, but this was not investigated within the present pipeline. Despite these exclusions, the 124 retained parameter sets provided a substantial ensemble for assessing parameter distributions and time-course prediction uncertainty under the imposed screening criteria.

\begin{figure}[t]
\centering

\newcommand{\parameterpanel}[2]{%
    \begin{minipage}[t]{0.44\columnwidth}
        \centering
        \includegraphics[width=\linewidth]{#1}
        \par\vspace{0.5mm}
        {\small\textbf{#2}}
    \end{minipage}%
}

\setlength{\tabcolsep}{1pt}

\begin{tabular}{cc}
\parameterpanel{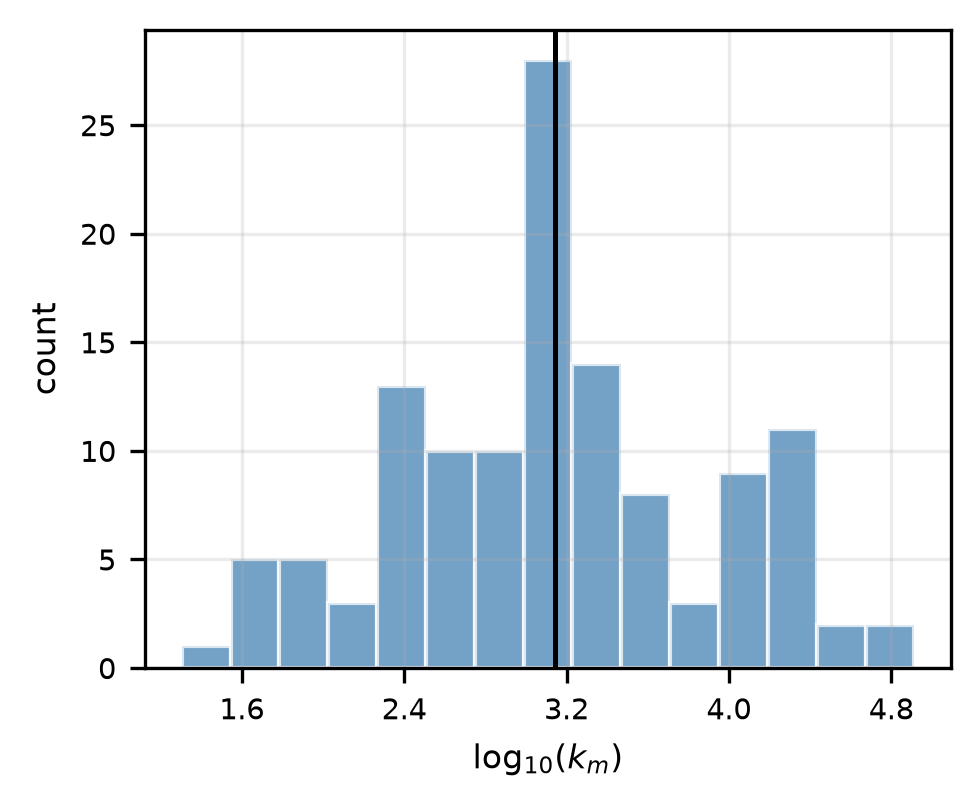}{(a)} &
\parameterpanel{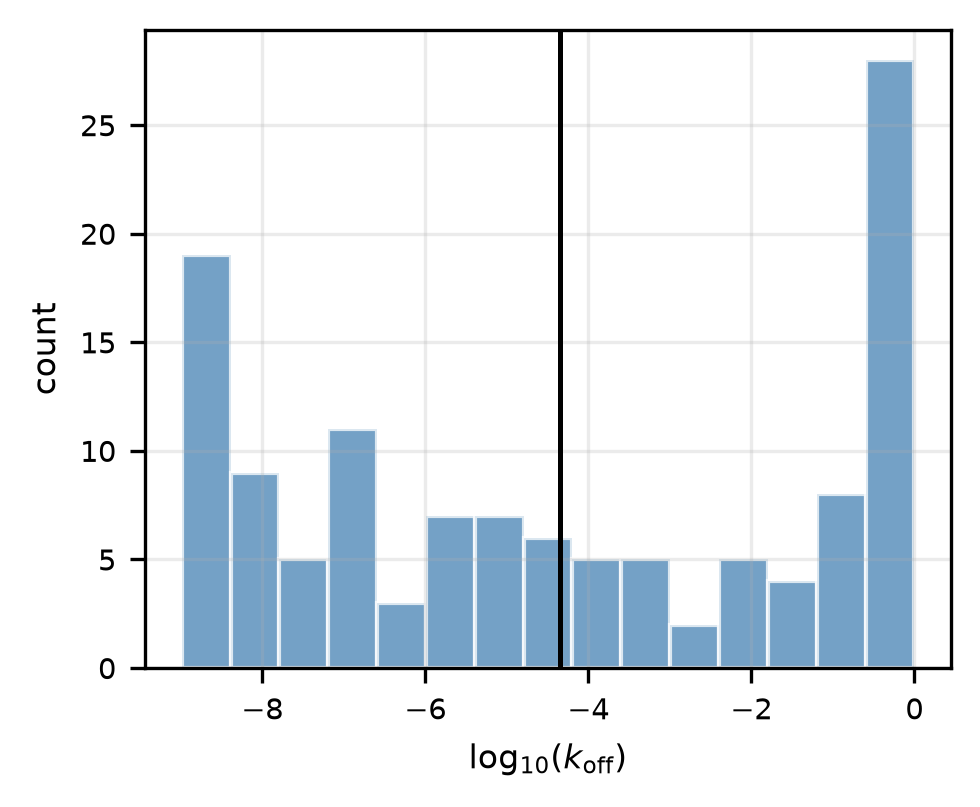}{(e)}
\\[1.5mm]
\parameterpanel{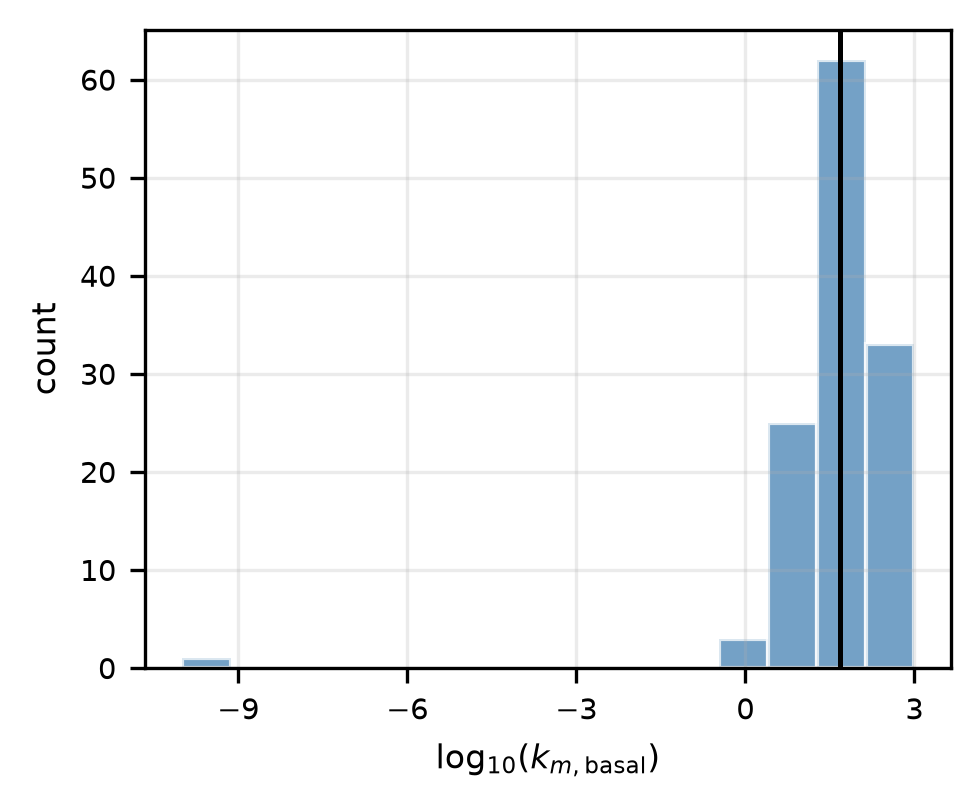}{(b)} &
\parameterpanel{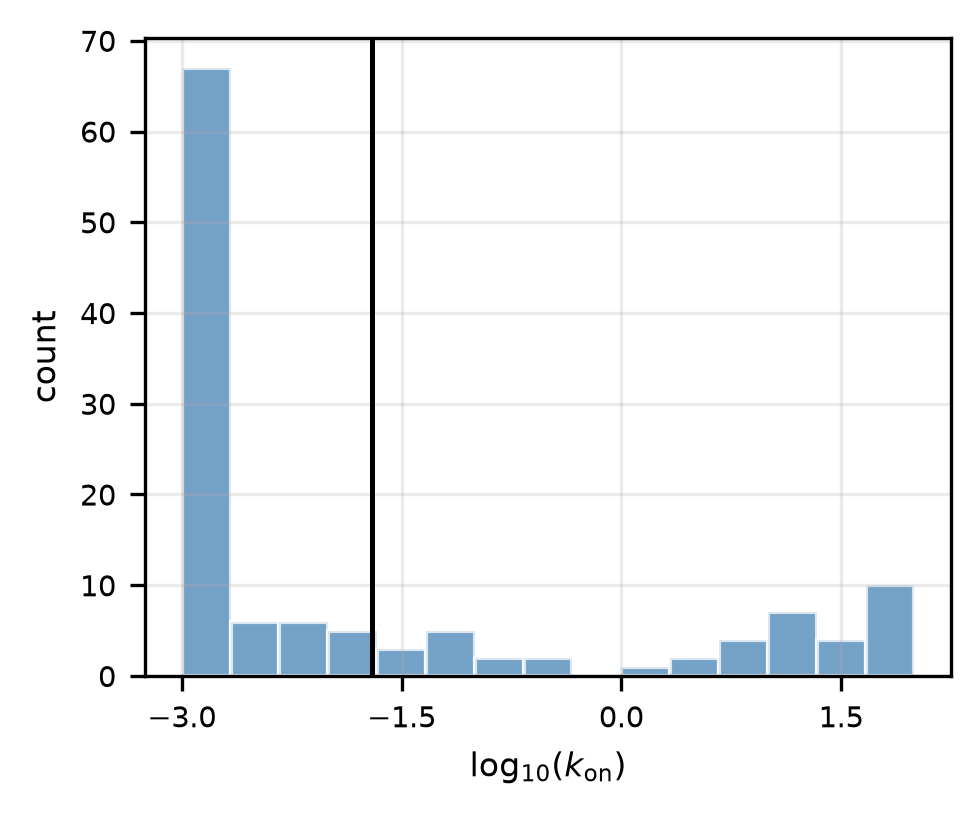}{(f)}
\\[1.5mm]
\parameterpanel{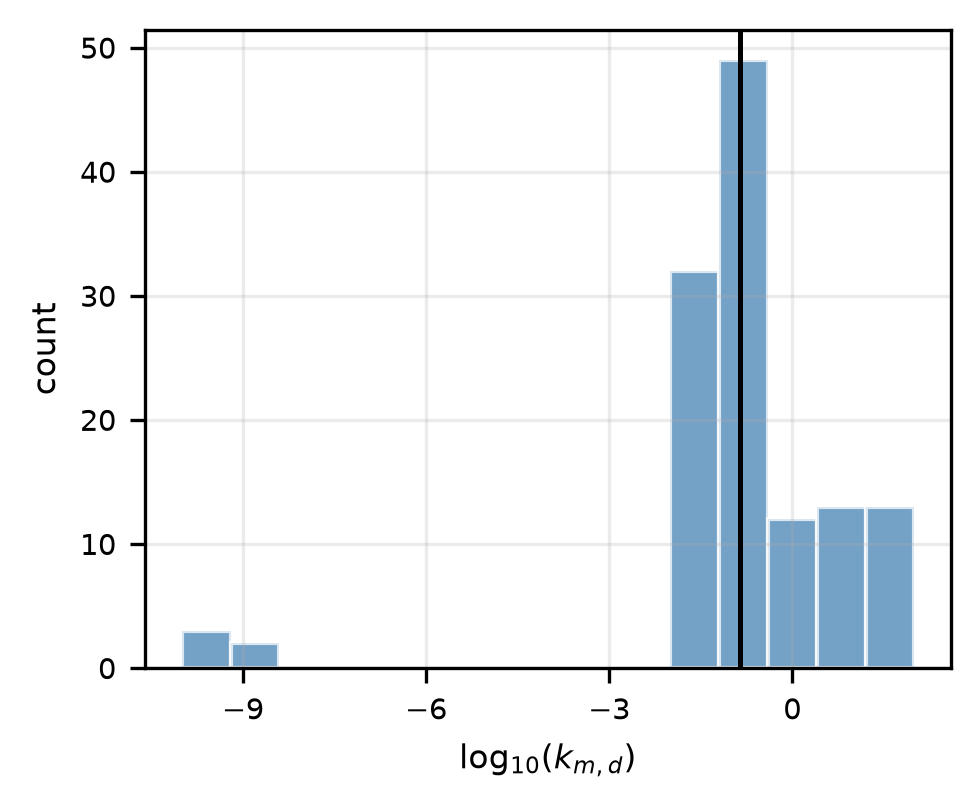}{(c)} &
\parameterpanel{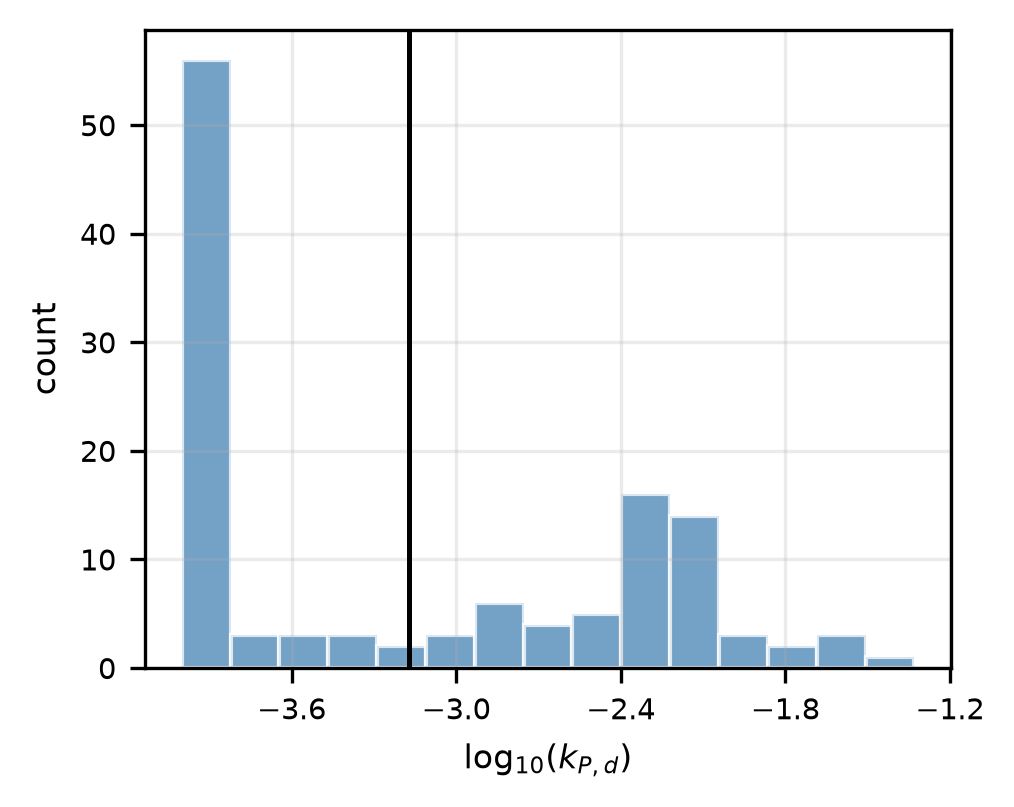}{(g)}
\\[1.5mm]
\parameterpanel{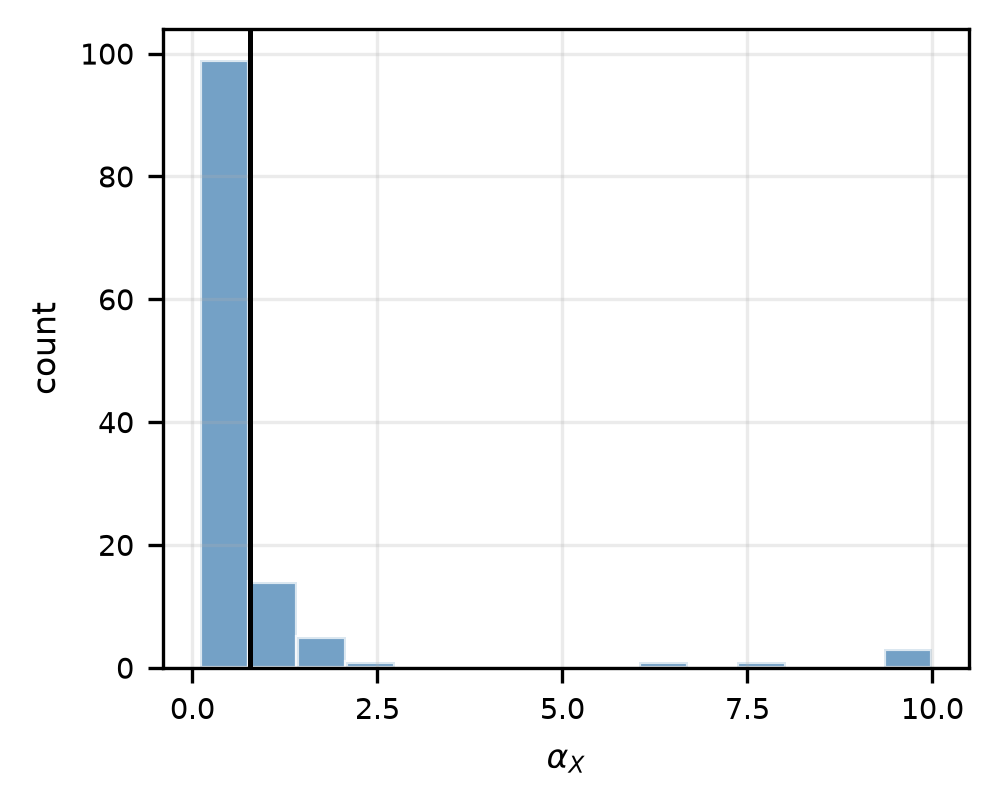}{(d)} &
\parameterpanel{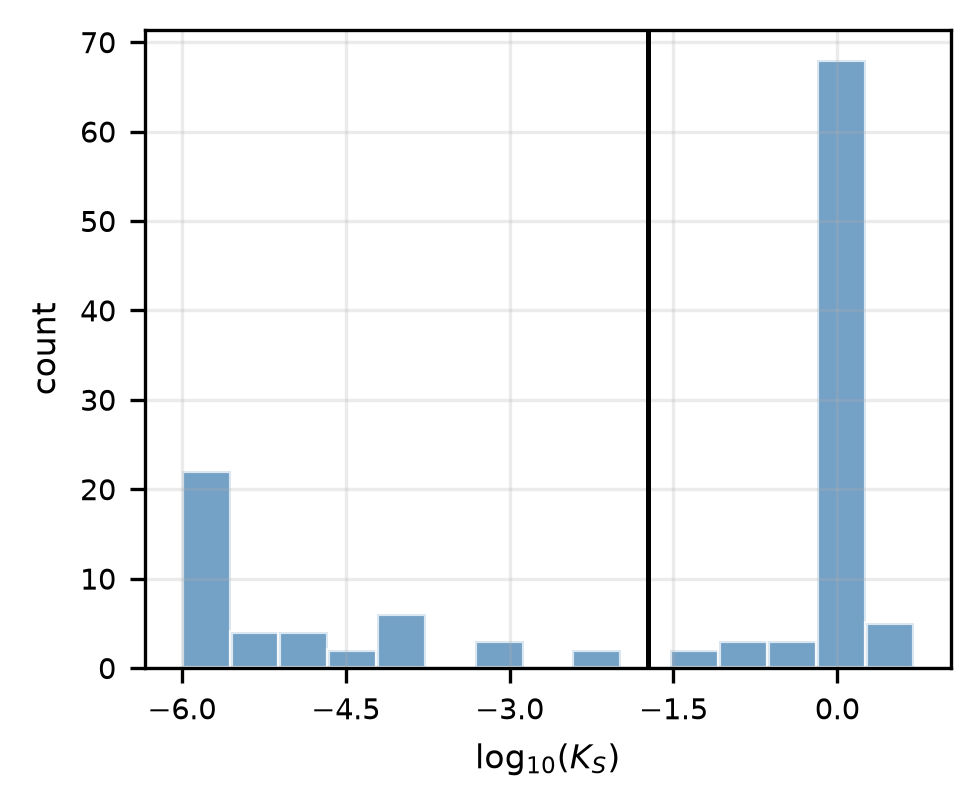}{(h)}
\end{tabular}

\caption{
Representative marginal distributions from the 23 fitted parameters in the
screened ensemble (124 parameter sets). The left column shows comparatively
concentrated distributions: (a) $k_m$, (b) $k_{m,\mathrm{basal}}$,
(c) $k_{m,d}$, and (d) $\alpha_X$. The right column shows broader
distributions: (e) $k_{\mathrm{off}}$, (f) $k_{\mathrm{on}}$,
(g) $k_{P,d}$, and (h) $K_S$. Parameters spanning several orders of magnitude are transformed using the base-10 logarithm. Vertical lines indicate ensemble means of log10-transformed or untransformed parameters, as appropriate.
}
\label{fig:parameter_distributions}
\end{figure}

Figure~\ref{fig:parameter_distributions} shows representative marginal parameter distributions from the screened parameter sets. For instance, the parameters in the left column, i.e., $k_m$, $k_{m,\mathrm{basal}}$, $k_{m,d}$, and $\alpha_X$, have relatively concentrated distributions around their respective ensemble means, showing that their estimates are more consistent across the retained models.. In contrast, the parameters in the right column, i.e., $k_{\mathrm{off}}$, $k_{\mathrm{on}}$, $k_{P,d}$, and $K_S$, have broader distributions, with some spanning several orders of magnitude. Logarithmic scales are used where appropriate to display these distributions.

These representative marginal distributions highlight differing degrees of parameter variability within the screened ensemble under the available data and assumed model structure. Although all retained models were fitted using the same data, model structure, and parameter bounds, and passed the same stability screening, differences remained in their parameter estimates. Satisfying the stability criterion therefore does not establish that the parameters were uniquely determined. Parameter identifiability remains a separate consideration \citep{villaverde2019identifiability,wieland_structural_2021}. Although the ensemble does not constitute a formal identifiability analysis, it still provides an empirical assessment of parameter uncertainty under the specified estimation and screening procedure.

\begin{figure}[!t]
\centering

\newcommand{\trajectorypanel}[1]{%
    \includegraphics[width=0.49\columnwidth]{#1}%
}
\setlength{\tabcolsep}{1pt}
\renewcommand{\arraystretch}{0.05}

\begin{tabular}{cc}
\trajectorypanel{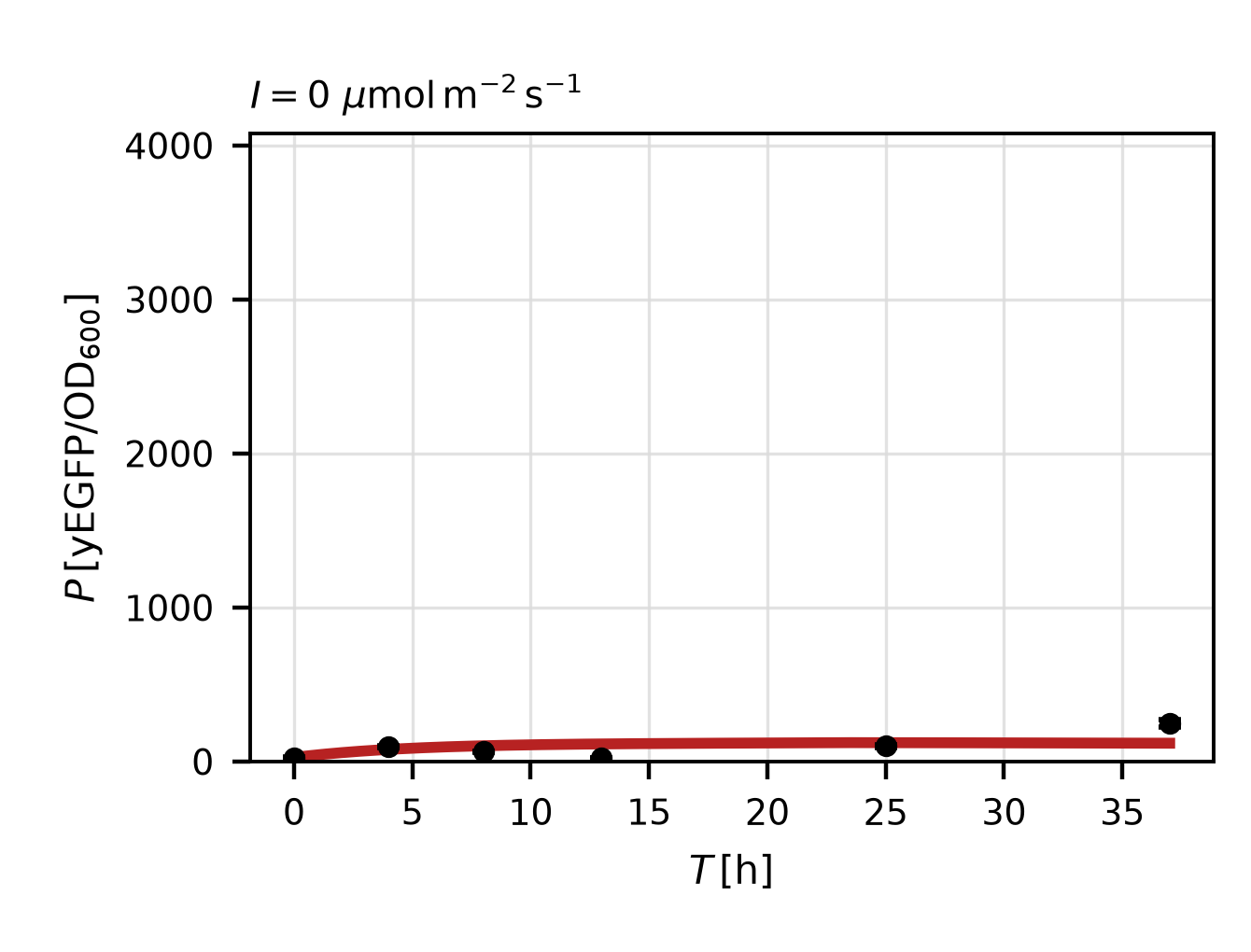} &
\trajectorypanel{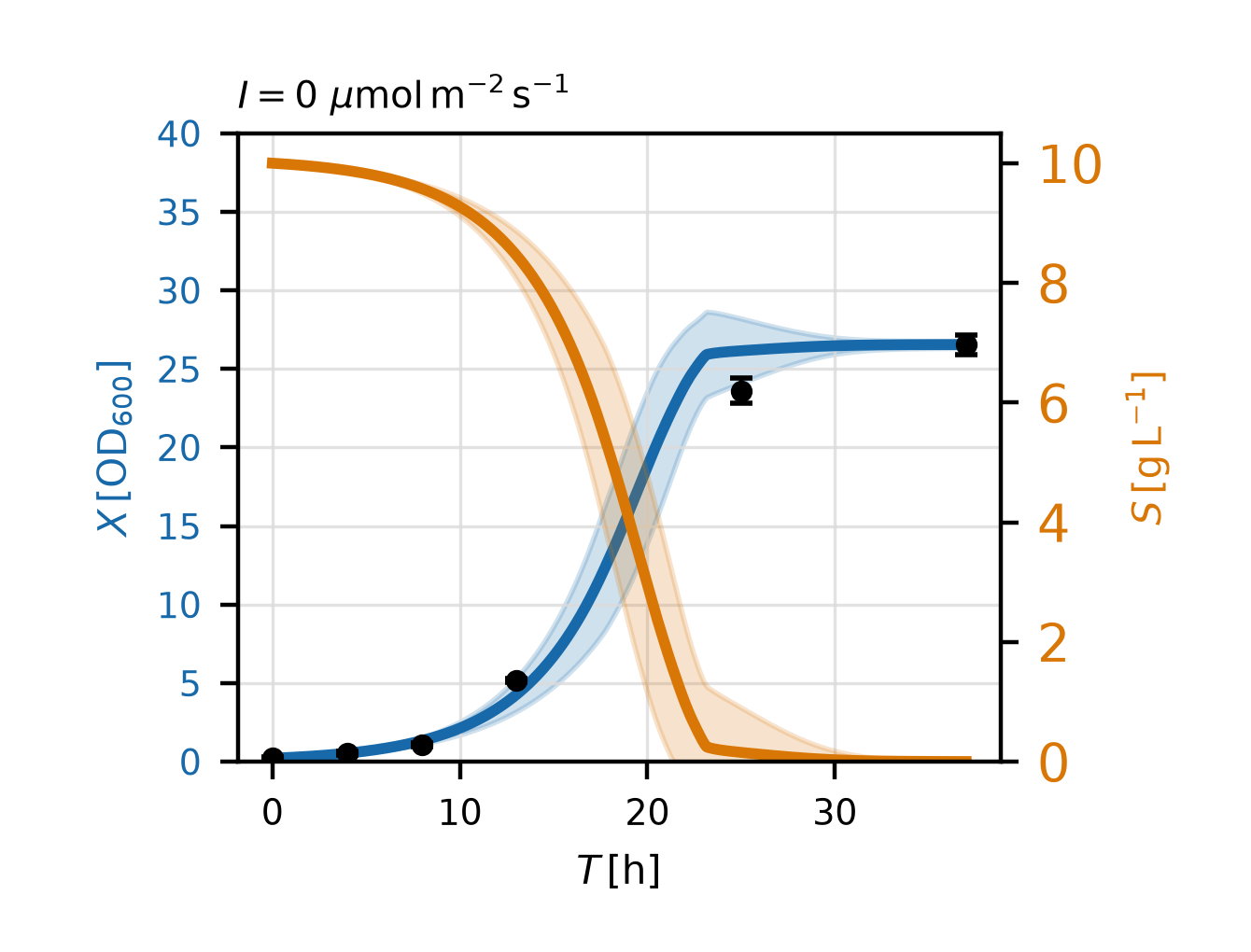}
\\[-1mm]
\trajectorypanel{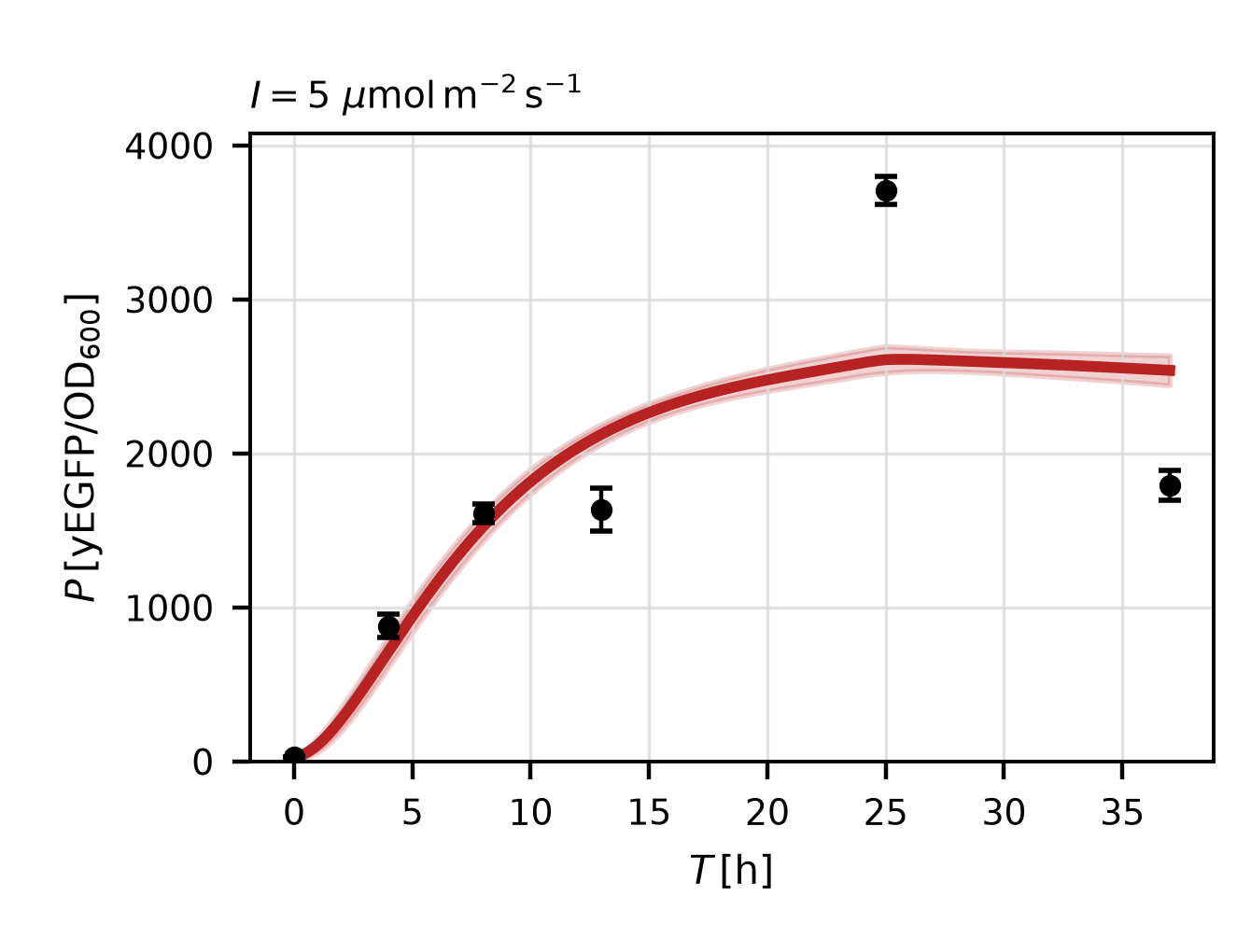} &
\trajectorypanel{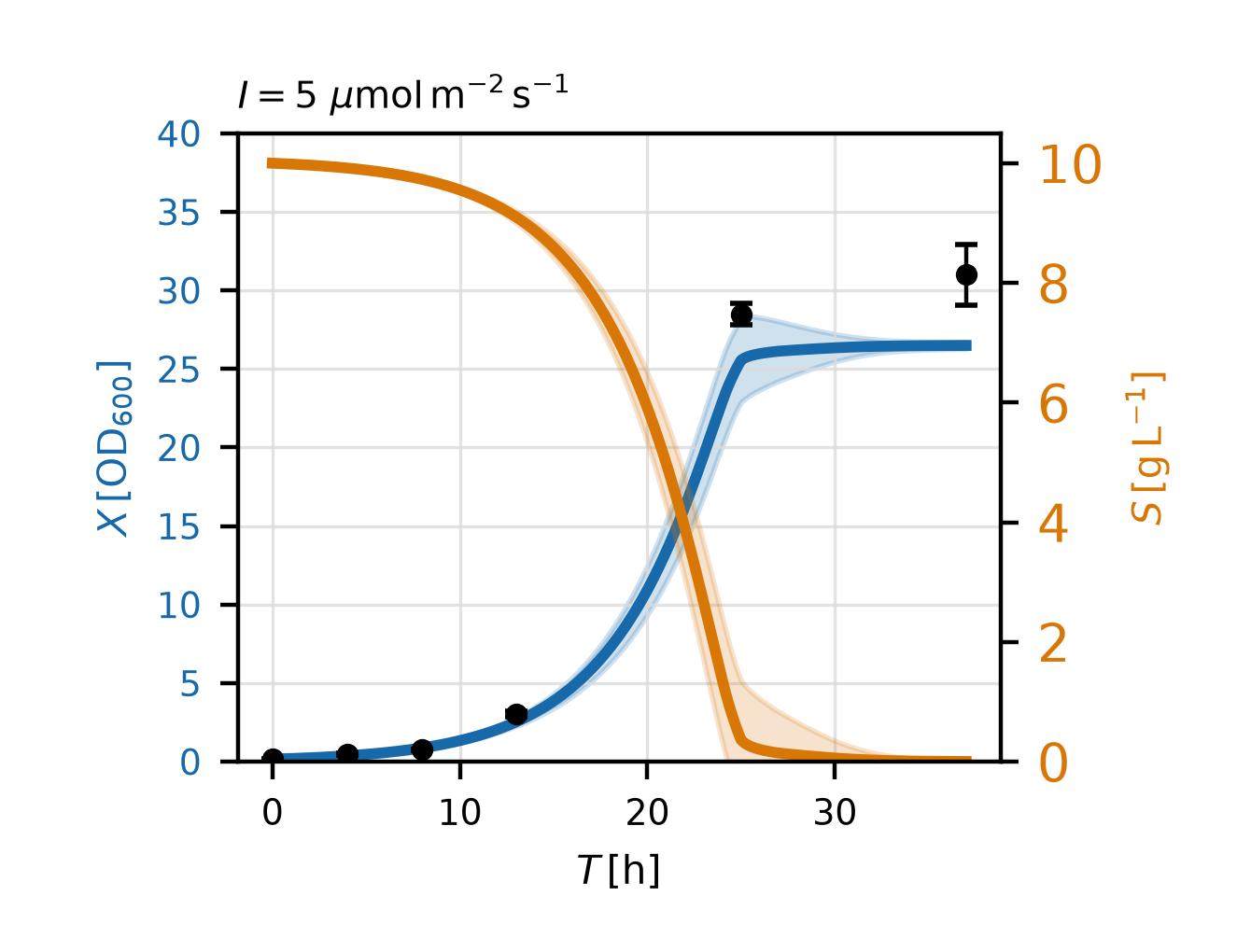}
\\[-1mm]
\trajectorypanel{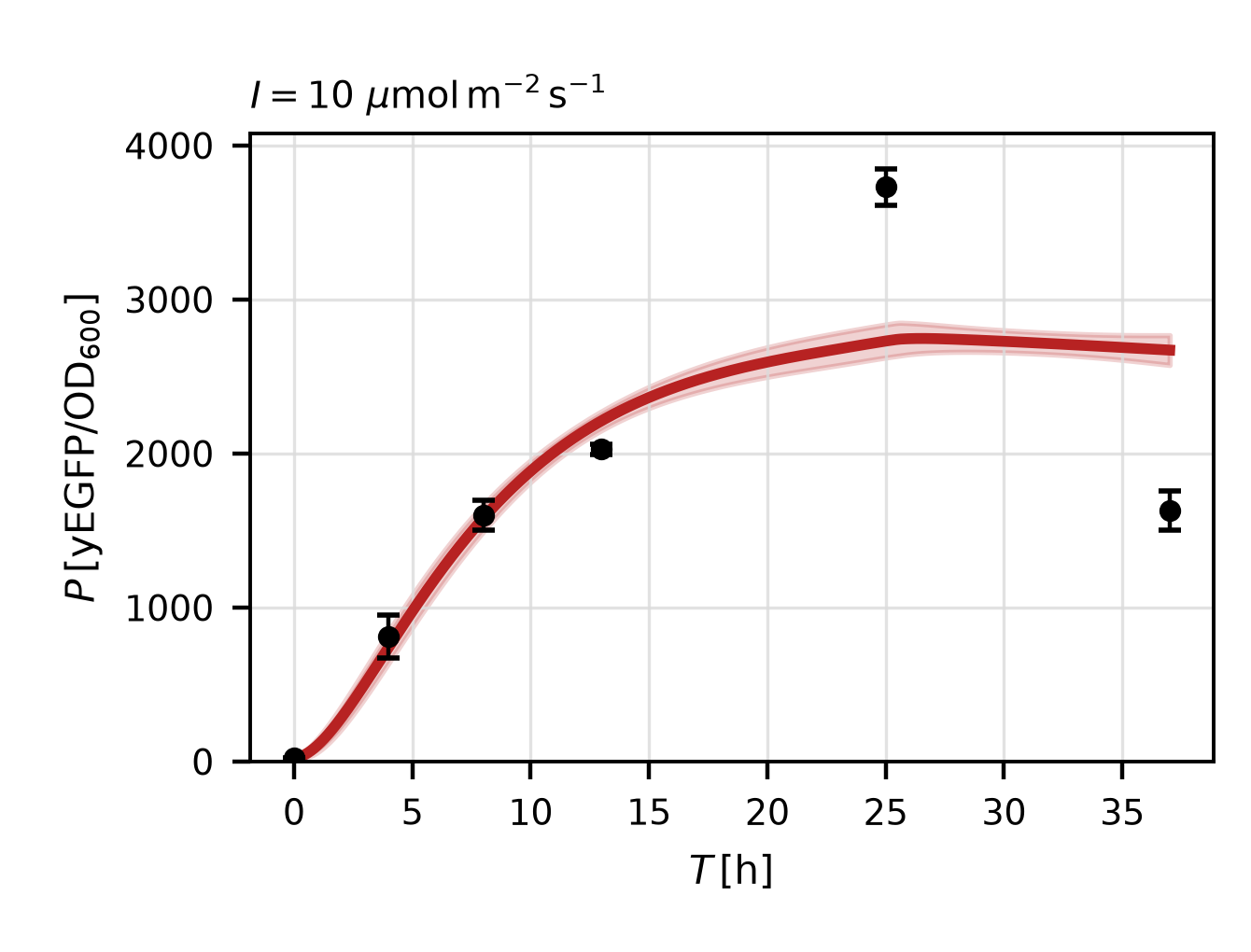} &
\trajectorypanel{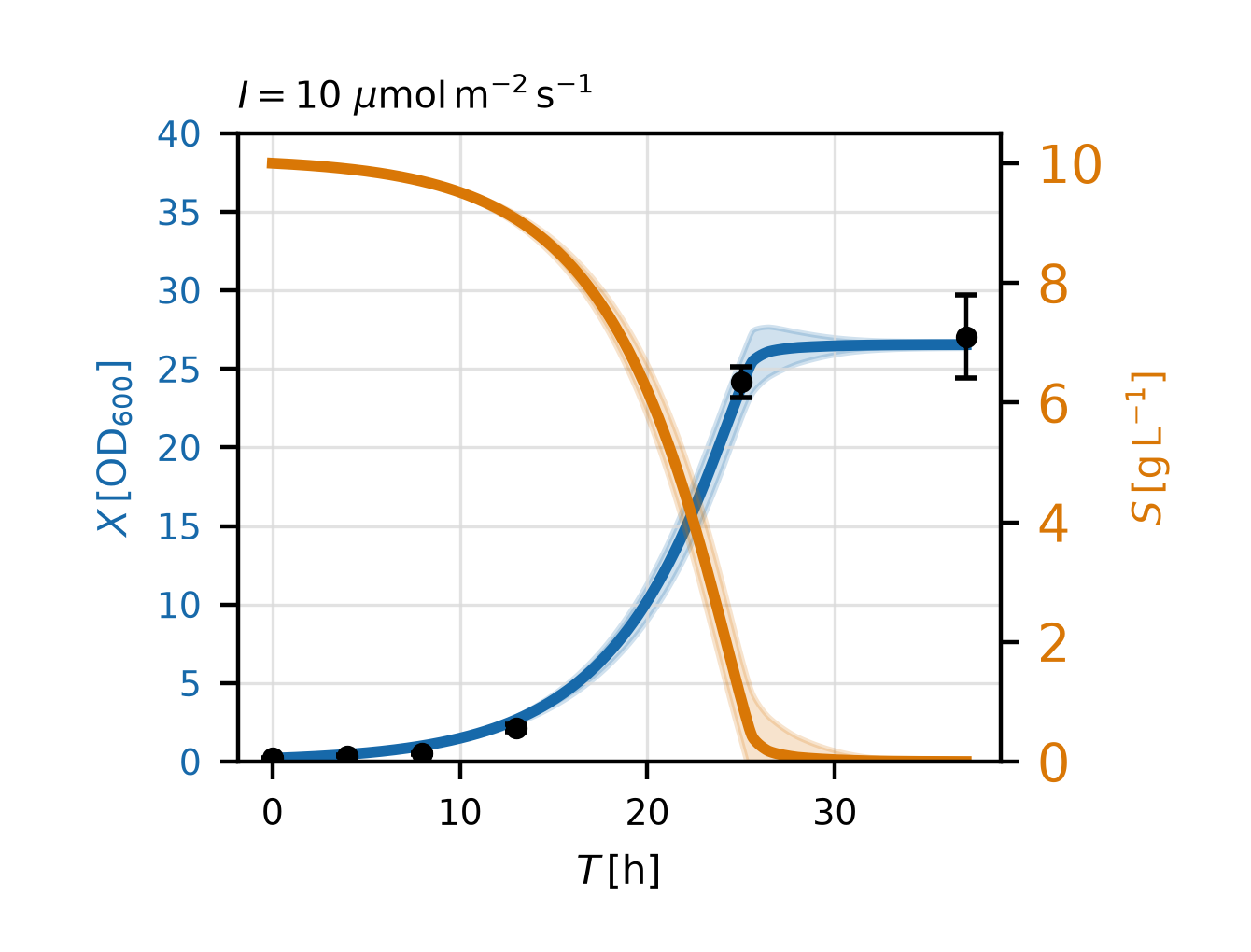}
\\[-1mm]
\trajectorypanel{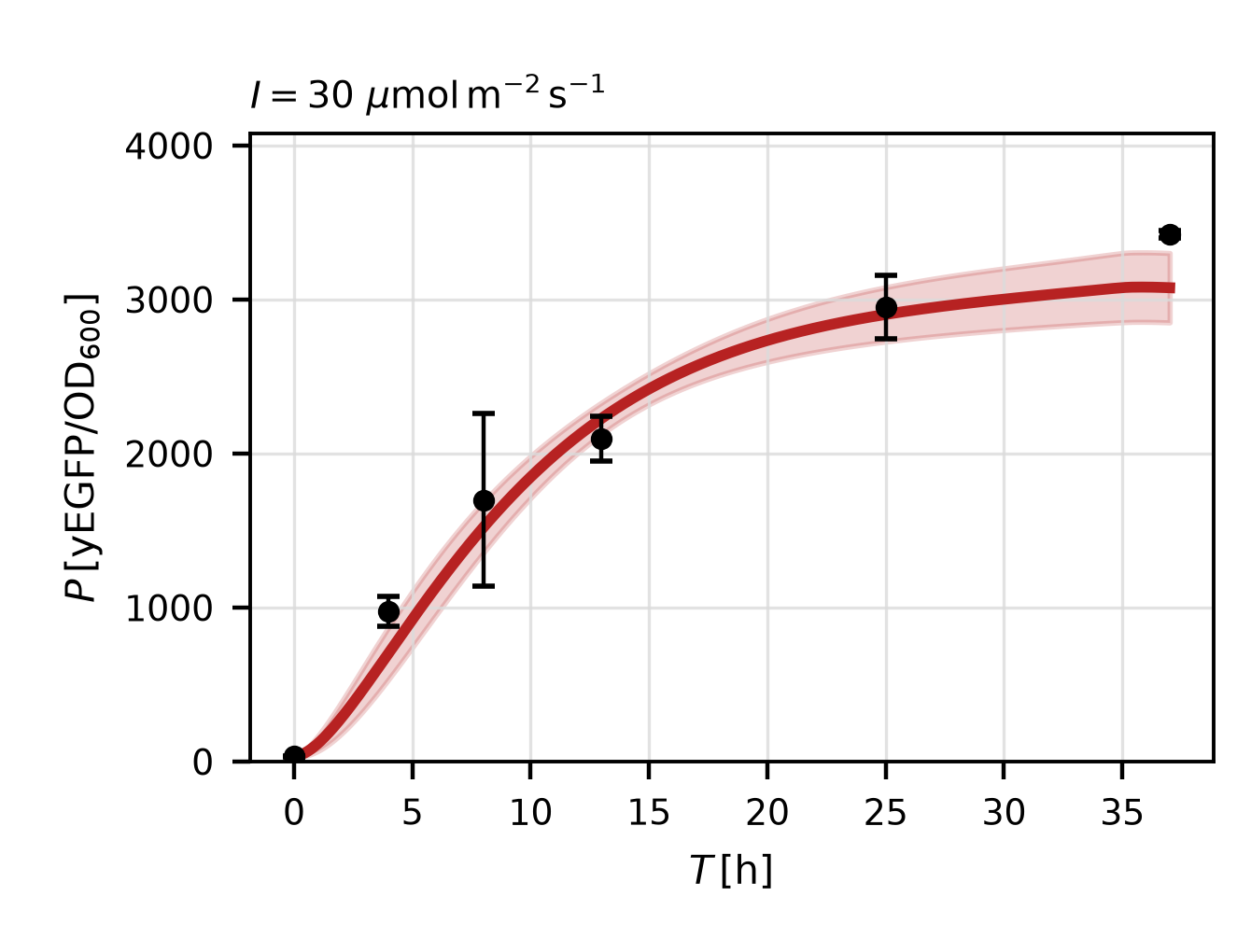} &
\trajectorypanel{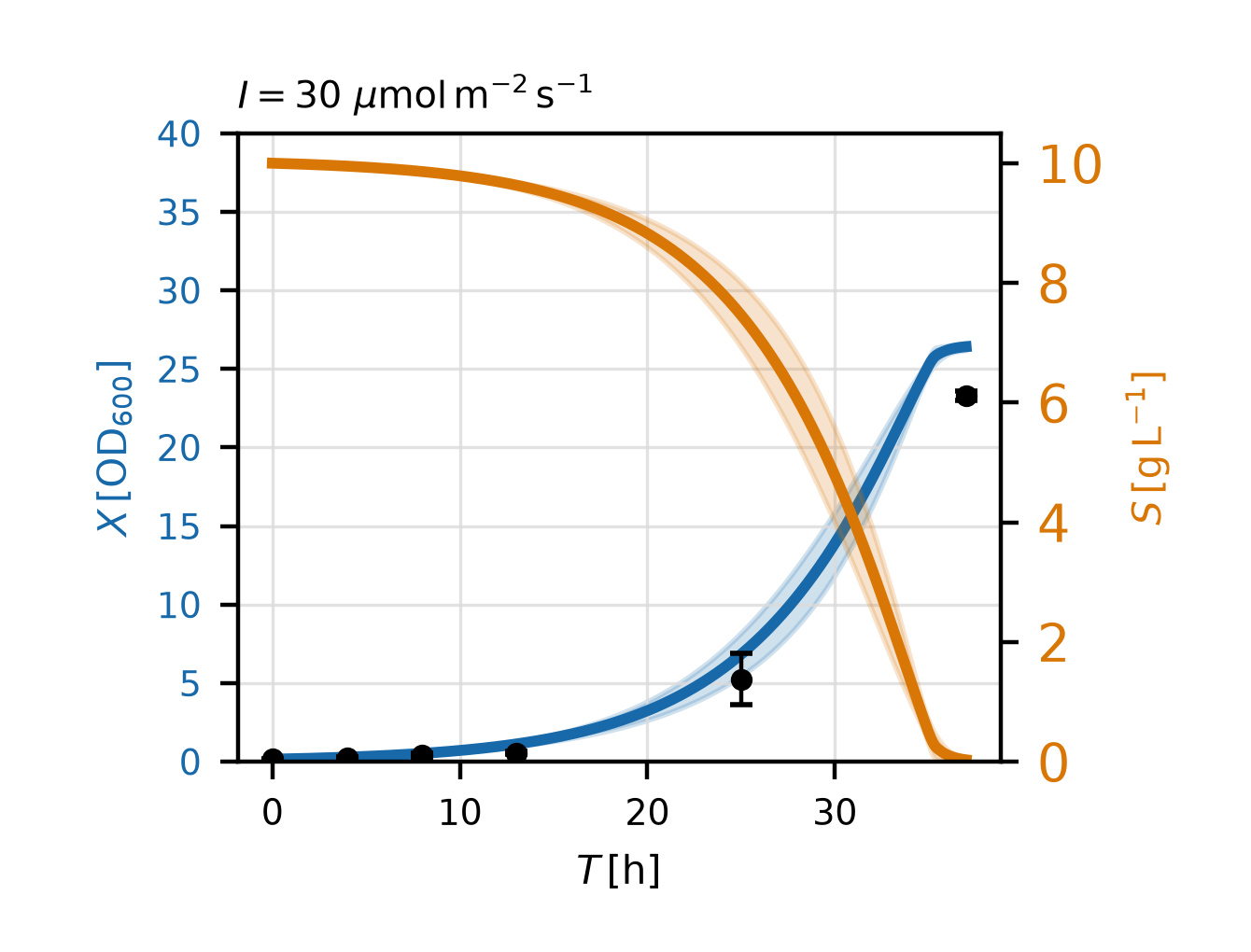}
\\[-1mm]
\trajectorypanel{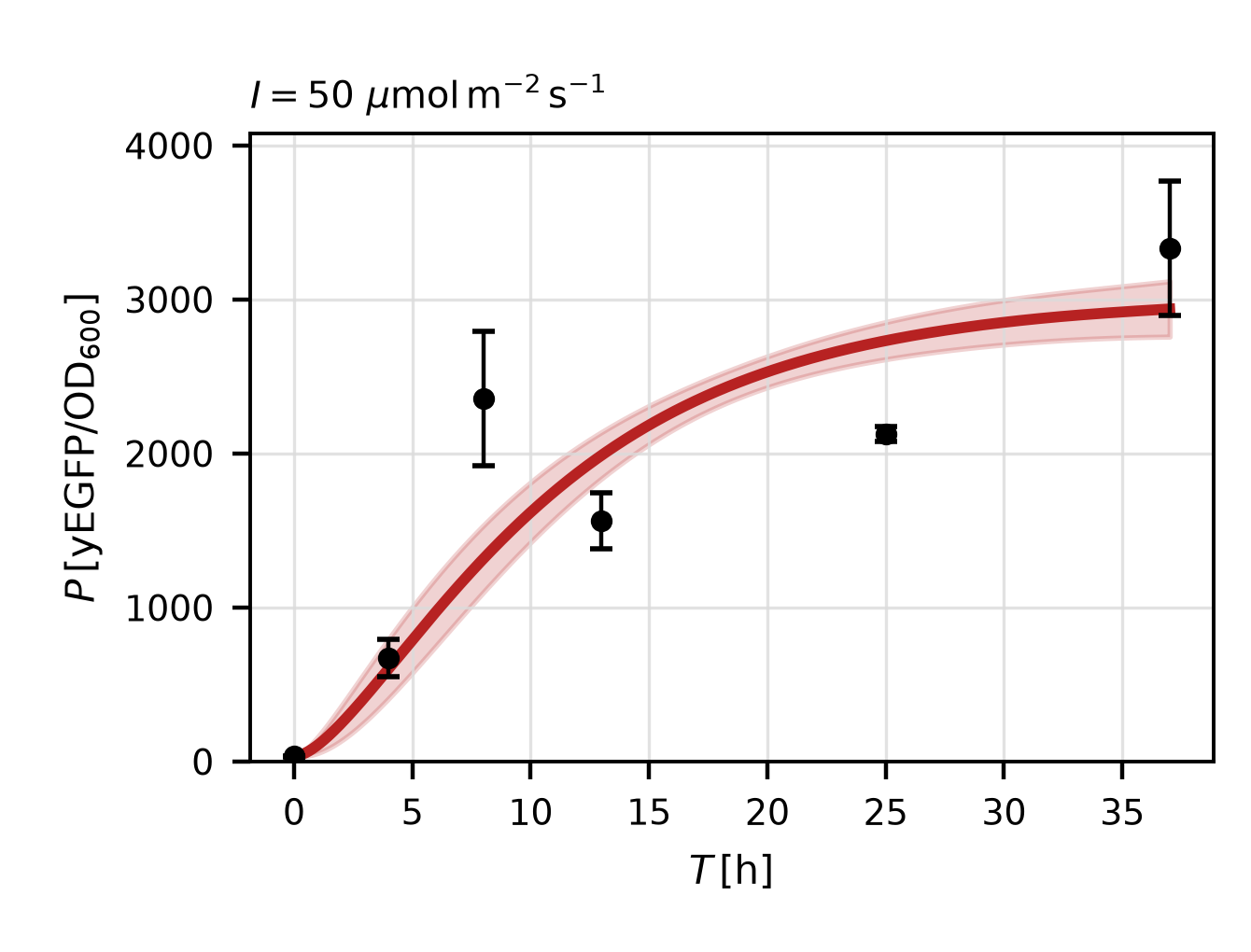} &
\trajectorypanel{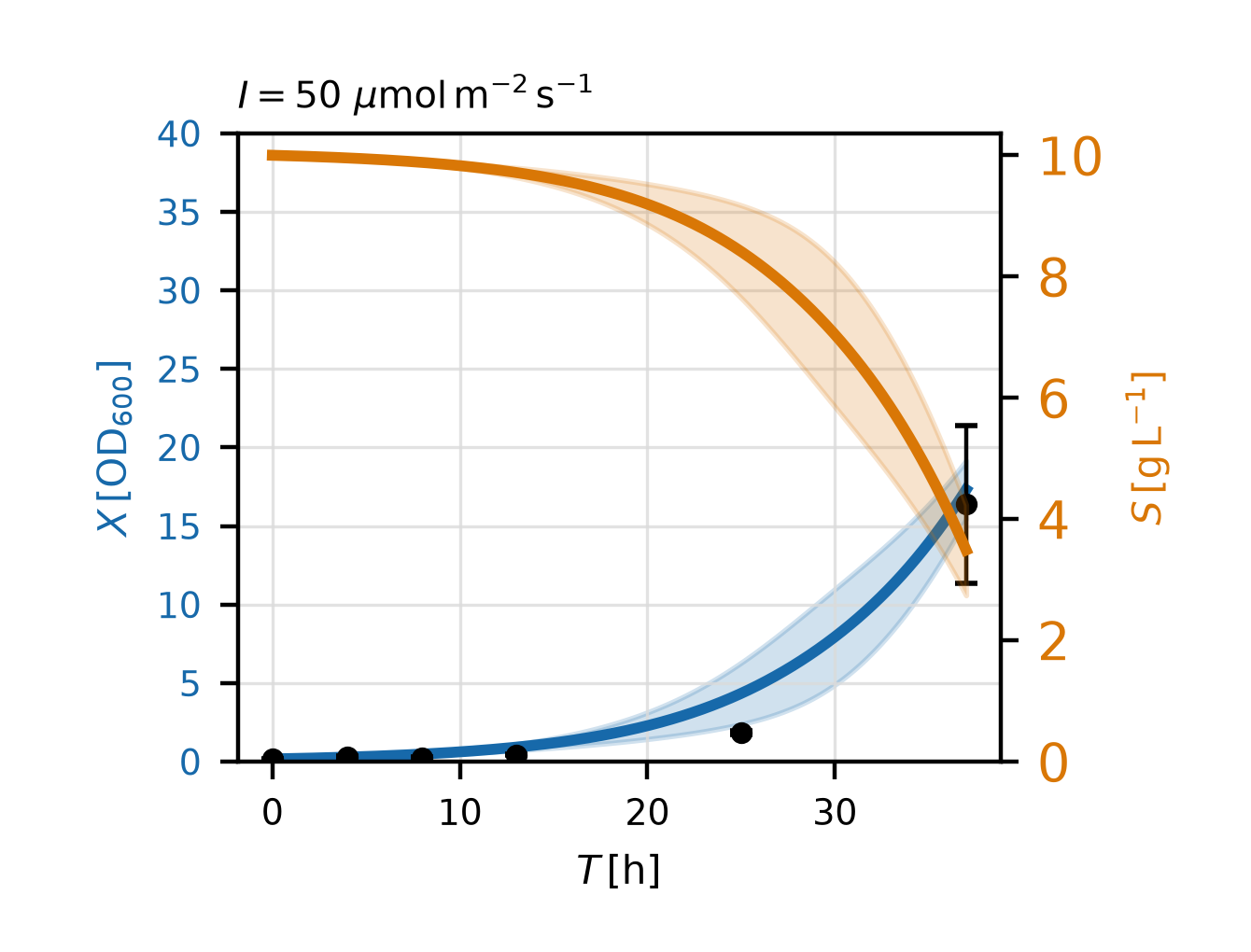}
\\[-1mm]
\trajectorypanel{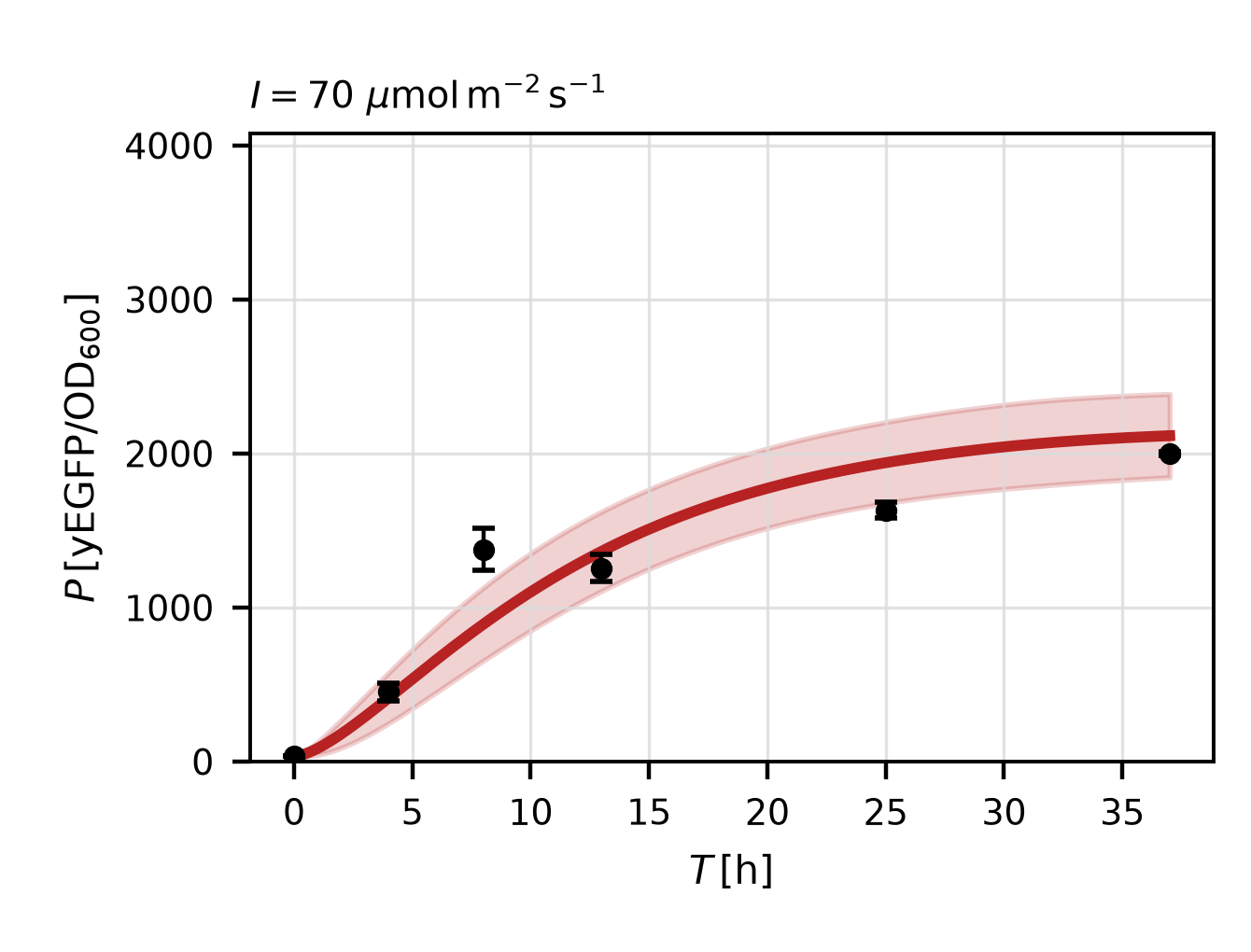} &
\trajectorypanel{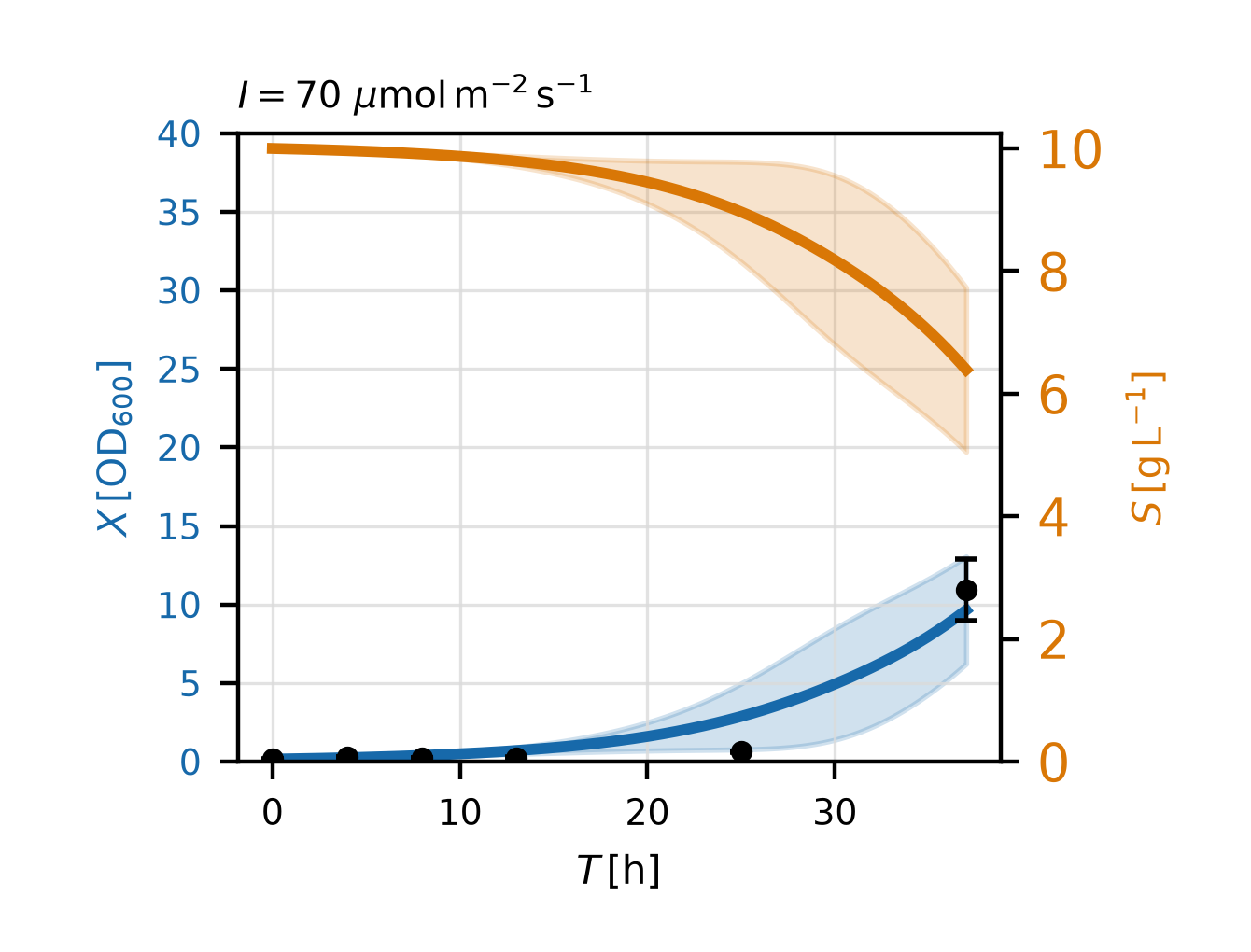}
\end{tabular}

\vspace{1mm}

\includegraphics[width=0.92\columnwidth]{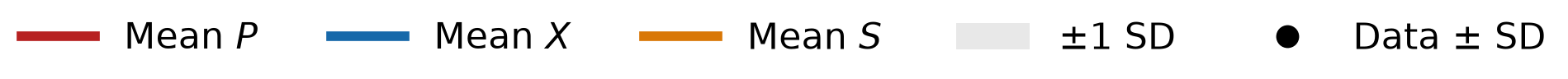}

\caption{
Ensemble predictions for protein $P$ (left) and biomass $X$ and substrate $S$ (right). Rows correspond to constant light intensities $I=0,5,10,30,50,70~\mu\mathrm{mol}\,\mathrm{m}^{-2}
\,\mathrm{s}^{-1}$, from top to bottom. Lines show means across the 124 retained models, with shading indicating $\pm$ one standard deviation. Markers and error bars show experimental
means $\pm$ one standard deviation from three independent biological replicates \citep{hoffman2025}.
}

\label{fig:ce8_trajectories}
\end{figure}

The full batch model was simulated using each of the 124 retained parameter sets under constant light intensities $I \in \{0,5,10,30,50,70\}\, \mu\mathrm{mol}\,\mathrm{m}^{-2}\,\mathrm{s}^{-1}$, matching the experimental conditions reported by \cite{hoffman2025}. Figure~\ref{fig:ce8_trajectories} shows the resulting time-course predictions for protein $P$, biomass $X$, and substrate $S$, together with the
experimental measurements for $P$ and $X$ \citep{hoffman2025}. Substrate predictions are shown for reference, as substrate concentration was not measured dynamically.

The ensemble-mean predictions captured the main trends in protein accumulation and cell growth observed experimentally. In darkness, the models predicted low yet nonzero protein production explained by basal expression, while biomass growth and substrate uptake were the fastest across the light conditions. As light intensity increased from 5 to $30~\mu\mathrm{mol}\,\mathrm{m}^{-2}\,\mathrm{s}^{-1}$, the maximum predicted protein accumulation increased, remaining similar at $50~\mu\mathrm{mol}\,\mathrm{m}^{-2}\,\mathrm{s}^{-1}$. At $70~\mu\mathrm{mol}\,\mathrm{m}^{-2}\,\mathrm{s}^{-1}$, the highest light intensity tested, predicted protein accumulation was markedly lower than at intermediate light intensities. Stronger illumination generally slowed biomass growth and predicted glucose depletion. Within the model, these responses reflect light-induced gene-expression dynamics together with stress-dependent photoinhibition that limits the rates of growth, transcription, and translation. Growth inhibition was already evident at intermediate light intensities ($\geq30~\mu\mathrm{mol} \, \mathrm{m}^{-2}\,\mathrm{s}^{-1}$), whereas a more marked reduction in predicted intracellular protein accumulation occurred at the highest intensity tested. These results underline the importance of balancing induction against photoinhibition when selecting control policies for optogenetic bioprocesses, with a view to improving productivity.

The widths of the shaded regions illustrate how parameter uncertainty propagates to the dynamic predictions. The ensemble-mean trajectories captured the overall trends, while the spread around them varied with the state variable and illumination condition. Prediction uncertainty was comparatively more apparent in the biomass and substrate trajectories at 50 and $70~\mu\mathrm{mol}\,\mathrm{m}^{-2}\,\mathrm{s}^{-1}$, and also in the protein predictions over the range $30–70~\mu\mathrm{mol}\,\mathrm{m}^{-2}\,\mathrm{s}^{-1}$. This may partly reflect the limited coverage of the growth dynamics under stronger illumination. Within the fixed experimental observation window, slower growth meant that later growth phases were less fully observed, providing less information to constrain the predicted dynamics.

Overall, compared with selecting a single fitted parameter set, this ensemble-based approach provides a broader representation of plausible model realisations supported by the available data, assumed model structure and screening procedure. The resulting empirical assessment of parameter-related epistemic uncertainty and its propagation to dynamic predictions can better inform optimisation and control strategies aimed at improving the efficiency of optogenetic bioprocesses.

\section{Conclusion}
This study integrated EL222-regulated optogenetic expression, light-associated stress, and macroscopic bioprocess dynamics into a compact mechanistic model. Ensemble-based parameter estimation was combined with \textit{a posteriori} screening for locally stable intracellular steady states, reflecting the assumed long-term intracellular behaviour under reference light conditions. This generated an ensemble of biologically plausible parameterisations consistent with the available experimental data, model structure, and screening criteria. The ensemble predictions captured the main experimental trends, while variability across parameter estimates and predicted trajectories highlighted the range of possible behaviours that would be overlooked by selecting a single fitted parameter set. Future work will focus on developing tailored model-based reinforcement learning and Bayesian optimisation approaches to optimise optogenetic protein production under phototoxicity. These approaches will explicitly account for parameter uncertainty in process optimisation and control-policy design.

\bibliography{ifacconf}

\end{document}